\documentclass[%
 reprint,
superscriptaddress,
nofootinbib,
 amsmath,amssymb,
pra,
longbibliography
]{revtex4-2}

\usepackage{dcolumn}
\usepackage{bm}
\usepackage{graphicx}
\usepackage{amsthm}
\usepackage{color}
\usepackage{xcolor}
\usepackage{verbatim}
\usepackage{esint}
\usepackage[english]{babel}
\usepackage{float}
\usepackage{amsfonts}
\usepackage{slashed}
\usepackage{latexsym}
\usepackage{bm}
\usepackage[colorlinks = true,
            linkcolor = blue,
            urlcolor  = blue,
            citecolor = blue,
            anchorcolor = red]{hyperref}
\usepackage{esint}
\usepackage{soul}
\usepackage{cancel}
\usepackage[normalem]{ulem}
\usepackage{empheq}
\usepackage{dsfont}
\usepackage{amsmath}
\usepackage{ulem}
\usepackage{epsfig}
\usepackage{tikz}
\usetikzlibrary{decorations.markings}
\def\la{\langle}
\def\ra{\rangle}

\def\be{\begin{equation}}
\def\ee{\end{equation}}

\begin{document}

\newcommand{\bigjprob}{{\mathcal{P}}}
\newcommand{\bigprob}{_{\bm{q}_F}{\mathcal{P}}_{\bm{q}_I}}
\newcommand{\cum}[1]{\llangle #1 \rrangle}       					
\newcommand{\op}[1]{\hat{\bm #1}}                					
\newcommand{\vop}[1]{\vec{\bm #1}}
\newcommand{\opt}[1]{\hat{\tilde{\bm #1}}}
\newcommand{\vopt}[1]{\vec{\tilde{\bm #1}}}
\newcommand{\td}[1]{\tilde{ #1}}
\newcommand{\mean}[1]{\la#1\ra}                  					
\newcommand{\cmean}[2]{ { }_{#1}\mean{#2}}       				
\newcommand{\pssmean}[1]{ { }_{\bm{q}_F}\mean{#1}_{\bm{q}_I}}
\newcommand{\ket}[1]{\vert#1\ra}                 					
\newcommand{\bra}[1]{\la#1\vert}                 					
\newcommand{\ipr}[2]{\left\la#1\mid#2\right\ra}            				
\newcommand{\opr}[2]{\ket{#1}\bra{#2}}           					
\newcommand{\pr}[1]{\opr{#1}{#1}}                					
\newcommand{\Tr}[1]{\text{Tr}(#1)}               					
\newcommand{\Trd}[1]{\text{Tr}_d(#1)}            					
\newcommand{\Trs}[1]{\text{Tr}_s(#1)}            					
\newcommand{\intd}[1]{\int \! \mathrm{d}#1 \,}
\newcommand{\stirlingii}{\genfrac{\{}{\}}{0pt}{}}
\newcommand{\dd}{\mathrm{d}}
\newcommand{\fullint}{\iint \! \mathcal{D}\mathcal{D} \,}
\newcommand{\drv}[1]{\frac{\delta}{\delta #1}}
\newcommand{\partl}[3]{ \frac{\partial^{#3}#1}{ \partial #2^{#3}} }		
\newcommand{\smpartl}[4]{ \left( \frac{\partial^{#3} #1}{ \partial #2^{#3}} \right)_{#4}}
\newcommand{\smpartlmix}[4]{\left( \frac{\partial^2 #1}{\partial #2 \partial #3 } \right)_{#4}}
\newcommand{\limit}[2]{\underset{#1 \rightarrow #2}{\text{lim}} \;}
\newcommand{\funcd}[2]{\frac{\delta #1}{\delta #2}}
\newcommand{\funcdiva}[3]{\frac{\delta #1[#2]}{\delta #2 (#3)}}
\newcommand{\funcdivb}[4]{\frac{\delta #1 (#2(#3))}{\delta #2 (#4)}}
\newcommand{\funcdivc}[3]{\frac{\delta #1}{\delta #2(#3)}}
\definecolor{dgreen}{RGB}{30,130,30}
\newcommand{\tkr}[1]{\textcolor{red}{#1}}
\newcommand{\ach}[1]{\textcolor{blue!70!red!80}{#1}}

\title{Supergrowth and sub-wavelength object imaging}
\author{Tathagata Karmakar}
	\email{tkarmaka@ur.rochester.edu}
	\affiliation{Department of Physics and Astronomy, University of Rochester, Rochester, NY 14627, USA}
	\affiliation{Center for Coherence and Quantum Optics, University of Rochester, Rochester, NY 14627, USA}
	\affiliation{Institute for Quantum Studies, Chapman University, Orange, CA 92866, USA}
	\author{Abhishek Chakraborty}
	\email{achakra9@ur.rochester.edu}
	\affiliation{Department of Physics and Astronomy, University of Rochester, Rochester, NY 14627, USA}
	\affiliation{Center for Coherence and Quantum Optics, University of Rochester, Rochester, NY 14627, USA}
	\affiliation{Institute for Quantum Studies, Chapman University, Orange, CA 92866, USA}
	\author{A.~Nick Vamivakas}
	\email{nick.vamivakas@rochester.edu}
	\affiliation{The Institute of Optics, University of Rochester, Rochester, NY 14627, USA}
	\affiliation{Materials Science, University of Rochester, Rochester, NY 14627, USA}
\author{Andrew N. Jordan}
\email{jordan@chapman.edu}
\affiliation{Institute for Quantum Studies, Chapman University, Orange, CA 92866, USA}
\affiliation{Department of Physics and Astronomy, University of Rochester, Rochester, NY 14627, USA}
\affiliation{Center for Coherence and Quantum Optics, University of Rochester, Rochester, NY 14627, USA}

\date{\today}

\begin{abstract}
We further develop the concept of supergrowth [Jordan,  Quantum Stud.: Math.~Found. \textbf{7}, 285--292 (2020)], a phenomenon complementary to superoscillation,  defined as the local amplitude growth rate of a function being higher than its largest wavenumber. We identify the superoscillating and supergrowing regions of a canonical oscillatory function and find the maximum values of local growth rate and wavenumber.  Next, we provide a quantitative comparison of lengths and relevant intensities between the superoscillating and the supergrowing regions of a canonical oscillatory function. Our analysis shows that the supergrowing regions contain intensities that are exponentially larger in terms of the highest local wavenumber compared to the superoscillating regions.   Finally, we prescribe methods to reconstruct a sub-wavelength object from the imaging data using both superoscillatory and supergrowing point spread functions.    Our investigation provides an experimentally preferable alternative to the superoscillation based superresolution schemes and is relevant to cutting-edge research in far-field sub-wavelength imaging.
\end{abstract}

\maketitle
\section{Introduction} \label{sec:intro}
 As we strive to engineer optical systems with better imaging capabilities, much attention has been devoted to surpassing the Rayleigh resolution limit in diffraction limited  optical devices \cite{goodmanIntroductionFourierOptics2017,lukosz_optical_1966}. To achieve superresolution, methods such as evanescent field based techniques \cite{pohl_optical_1984,yang_super-resolution_2014} and  negative refractive index materials \cite{PhysRevLett.85.3966} have been proposed. To that end, the efficacy of superoscillatory spot generation for  label free far-field based superresolution imaging is well established. Superoscillation (SO) refers to the phenomena of local oscillation frequency of a function being faster than its fastest Fourier component \cite{Aharonov-preprint,ferreiraSuperoscillationsFasterNyquist2006,kempfBlackHolesBandwidths2000,kempfFourAspectsSuperoscillations2018,tangScalingpropertiessuperoscillations2016, chen_superoscillation_2019}. Thus, in optical systems, superoscillation   can be utilized to generate sub-wavelength hot-spots and thus beat the Rayleigh resolution limit \cite{Berry2006May,berry2019roadmap,rogers_optical_2013}. Sub-wavelength spots have been realized at the focal plane of a microscope objective using optical eigenmode approach implemented with a spatial light modulator \cite{baumgartl_far_2011}. Also, superoscillatory spot properties of radially polarized Laguerre-Gaussian beams in a confocal laser scanning microscopy setup have been studied  \cite{kozawa_numerical_2015,Kozawa2018Feb}. Optimization of superoscillatory lenses for sub-diffraction limit optical needle generation has been investigated as well \cite{diao_controllable_2016}.
 
 In principle, it is possible to design arbitrarily small superoscillatory optical spots. However, superoscillation is necessarily accompanied by enhanced side lobes. This leads to poor quality in imaging and unrealistic constraints on the dynamic ranges of the detectors. Several recent ventures try to address this problem through numerical or design based approaches.  Simulations for simultaneous optimization of superoscillatory spot size and their relative intensities compared to the side lobes have been performed \cite{rogersOptimisingSuperoscillatorySpots2018}. Elimination of sidelobes along a particular dimension  by introducing moonlike apertures has been demonstrated  \cite{hu_optical_2021}. 
 
 In this work, we show how the problem of enhanced sidelobes can be circumnavigated by utilizing supergrowing functions, a concept proposed by Jordan \cite{jordanSuperresolutionUsingSupergrowth2020}. The phenomenon of supergrowth (SG)  is analogous to superoscillation.  While superoscillation pertains to the local oscillation rate of a function, supergrowth occurs when the local growth rate of the amplitude is higher than the highest wave number in the Fourier space of a bandlimited function. A large local growth rate enables enhanced spatial resolution. The idea is analogous to evanescent wave imaging microscopes, but is applicable in the far-field. In our analysis, we consider a canonical single-parameter bandlimited oscillatory function and locate its SO and SG regions. In addition to the entire SO and SG areas, we identify the near maxima regions for both phenomena. We provide analytical estimates for the intensities in these regions and find that the amount of light in the SG areas is exponentially  higher compared to the regions associated with SO. Finally, we present two parallel schemes to reconstruct an incoherently illuminated  sub-wavelength object   with  SO  and SG point spread functions (PSF). We numerically compare our approaches to object reconstruction using a bandlimited $sinc$ PSF. As expected,  superresolved object reconstruction is achieved in the former two cases.

 The past two decades have witnessed considerable progress in the superoscillation related research.   The far-field nature of superoscillatory fields have been shown by proving that subwavelength structures generated by a diffraction grating can survive farther than the evanescent waves \cite{berryEvolutionQuantumSuperoscillations2006}. This investigation has inspired further studies into Schr\"{o}dinger equation based evolution of superoscillatory waves \cite{ 117095701,138046077,berryEscapingSuperoscillations2018}. The correspondence between superoscillations and weak values  \cite{294456787,182010239} is well known. On a related note, super-phenomena in arbitrary quantum observables \cite{jordanSuperphenomenaArbitraryQuantum2022} have been proposed. Towards the more implementational side,  numerical optimization of the energy ratio between superoscillatory region and total signal has been studied \cite{36243494}.   Also new methods for generating superoscillatory functions have been proposed \cite{7203461,smithOpticalvorticescoherence2019,chojnacki_new_2016}.  In comparison supergrowth is a very recent concept. Jordan \cite{jordanSuperresolutionUsingSupergrowth2020} showed that it is possible to access superresolving features using supergrowth. Spherical Bessel function based method for systemic generation of SO/SG functions and general approximation scheme using bandlimited functions have been prescribed \cite{karmakar2023bessel}.  Our analysis draws inspiration from these works and solidifies the benefits of implementing supergrowth based superresolution imaging in practice.

This article is organized as follows. In Sec.~\ref{so-sg-prop}, we briefly describe the phenomena of SO and SG  using a canonical oscillatory function. Sec.~\ref{so-sg-propa}   compares the length of SO and SG regions for the chosen function and Sec.~\ref{so-sg-intensity} compares their intensities. In Sec.~\ref{so-sg-imaging}, we present the schemes to reconstruct a sub-wavelength object using both superoscillatory and supergrowing spots. In Sec.~\ref{disc} we discuss the implications of our findings. We conclude in Sec.~\ref{concl}. 

\section{\label{so-sg-prop} Properties of SO/SG functions}
For our analysis, we consider the function
\cite{berryEvolutionQuantumSuperoscillations2006}
\begin{equation}
    f(x)=(\cos x+i a\sin x)^N,
    \label{f1}
\end{equation}
parameterized by a positive real number $a$, which sets its SO and SG properties, and a natural number $N$, which gives an upper bound to the Fourier wavenumber $k_{\rm max}=N$ for the function. This corresponds to a shortest wavelength of oscillation of $\lambda_{\rm min} =2\pi/N$. In the following sections we characterize the SO and SG regions of $f(x)$ and look at the intensities within.
\subsection{\label{so-sg-propa} SO/SG regions}
\begin{figure}
    \centering
    \includegraphics[width=\linewidth]{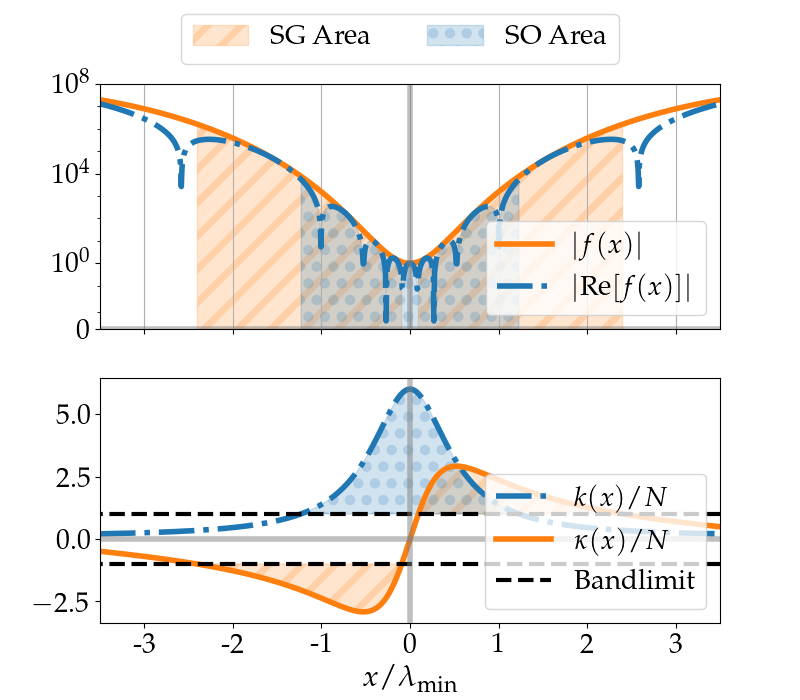}
        \caption{The solid orange (blue dot-dashed) curve in the upper panel shows the magnitude (real part) of the function defined in Eq.~\eqref{f1} in log scale for $a=6$ and $N=10$. The vertical grid in the top panel has a spacing equal to the smallest wavelength allowed by the bandlimit. Clearly near $x=0$  the function oscillates on a scale smaller than the smallest wavelength possible. This is  also apparent from the bottom panel where the  the local rates of oscillation from Eq.~\eqref{k} (blue dot-dashed line) and growth from  Eq.~\eqref{kappa} (solid orange line) and  are shown as functions of position.  The black dashed lines indicate the bandlimit. In both panels, the shaded regions indicate where the function is superoscillating (blue dots) and supergrowing (orange hatching).}
    \label{fns}
\end{figure}

For any complex valued function $f(x)$, the local wavenumber is defined as $k(x) = \text{Im }\partial_x\log f(x)$, and the local growth rate is $\kappa(x)=\text{Re }\partial_x\log f(x)$. If $f(x)$ is bandlimited with highest wavenumber $k_{\text{max}}$, the function is said to be  superoscillating (supergrowing) at $x$ if $|k(x)|>k_{\rm max}$ ($|\kappa(x)|>k_{\rm max}$).\par

For $f(x)$ in Eq.~\eqref{f1}, local wavenumber and growth rate take the form \cite{jordanSuperresolutionUsingSupergrowth2020}
\be
k(x) = N  \frac{a}{\cos^2 x + a^2 \sin^2 x},
\label{k}
\ee
and
\be
\kappa(x) = N \frac{(a^2-1) \sin x \cos x}{\cos^2 x + a^2 \sin^2 x}.
\label{kappa}
\ee
For  $\sqrt{2}-1<a<\sqrt{2}+1$, this function shows SO, but there is no SG. For $a>\sqrt{2}+1$, the function displays both SO and SG behavior. We will show in Sec.~\ref{so-sg-intensity} that larger values of $a$ show more SG but less SO, so we will consider only such cases.

Fig.~\ref{fns} shows the behavior of the function in Eq.~\eqref{f1} (top panel) and the local rates of oscillation and growth in Eqs.~\eqref{k} and \eqref{kappa} (bottom panel) for $a=6$ and $N=10$. The function is superoscillating ($k(x)\ge N$) in $x\in \left[\arctan  (1/\sqrt{a}), \arctan  (-1/\sqrt{a})\right]$, shaded orange in Fig.~\ref{fns} (bottom panel). The largest local wavenumber is $k(x=0) = a N$, and it is clear (from top panel) that near origin the function oscillates on a scale shorter than the shortest wavelength. 
  
The function shows SG behavior ($\kappa(x)\ge N$) in $x \in [\arctan y_l,\arctan y_h]$, with 
\begin{equation}
    \begin{split}
        &y_l=\tfrac{1}{2}\left(1-\tfrac{1}{a^2}-\sqrt{(1-\tfrac{1}{a^2})^2-\tfrac{4}{a^2}}\right),\\
        &y_h=\tfrac{1}{2}\left(1-\tfrac{1}{a^2}+\sqrt{(1-\tfrac{1}{a^2})^2-\tfrac{4}{a^2}}\right).
    \end{split}
    \label{ylyhtxt}
\end{equation}
 The largest growth rate occurs when $ x = \pm x_{sg}= \pm \arctan(a^{-1})$, giving a maximum growth rate of
\be
\kappa_{\max} = \pm \frac{N}{2} (a - a^{-1}).
\ee
For large values of $a$, this approaches $\pm N a/2$ rapidly.

\subsection{\label{so-sg-intensity}Intensity comparison}
In imaging applications, SO/SG spots with higher intensities are ideal since that makes superresolution imaging less susceptible to the influence from sidelobes. In this section, we quantify the amount of light in  $f(x)$ in regions showing SO/SG behaviors. For simplicity, we consider the function $f(x)$ to be the intensity point spread function of an incoherent imaging device \cite{wilson_resolution_2011}. The total intensity  within a single period of $f(x)$ can be defined as
\be
I_0 = \int_{-\pi}^{\pi} dx |f(x)|^2.
\label{I0}
\ee
For large $a$ and $N$, the total intensity can be approximated as (see Appendix.~\ref{app0})
\be
I_0\approx 2\pi\frac{(2N)!}{(N!)^2}\left(\frac{a}{2}\right)^{2N}.
\label{Itot_approx}
\ee
We can further simplify the expression above using Stirling's approximation for factorials
\be
I_0\approx 2\sqrt{\frac{\pi}{N}}a^{2N}.
\label{Itot_approx_sl}
\ee
We see indeed that $I_0$ increases exponentially in $N$. From Fig.~\ref{fns}, it is clear that the most of this light is available away from the origin. This is a manifestation of  pronounced side-lobes around a superoscillatory spot. 
However, as the subsequent analysis will reveal, SG based imaging can be highly advantageous in this respect. 

We see at the superoscillatory region $x=0$, while the region of superoscillation (where $k > N$) goes from $x \in [-\arctan\tfrac{1}{\sqrt{a}}, \arctan\tfrac{1}{\sqrt{a}}]$, 
the useful range is more restrictive.  This is where the oscillations are at their fastest, and the amplitude of oscillations is approximately constant.  Noting that $\ln f \approx i a N x + \tfrac{1}{2} N (a^2-1) x^2$, near $x=0$, we get a more restrictive range $x \in [-1/\sqrt{N(a^2-1)}, 1/\sqrt{N(a^2-1)}]$. We can find the amount of power in the range by integrating $|f|^2$ over this restrictive interval to find (Appendix.~\ref{app0}),
\be
I_{\mathrm{SO,R}} \approx
\frac{2e^{1/4}}{\sqrt{N(a^2-1)}}.
\label{Iso_restr}
\ee
We note that the intensity in this region is exponentially suppressed compared to the total intensity in Eqs.~\eqref{Itot_approx},\eqref{Itot_approx_sl}.  

Taking a more relaxed view and using the whole region of superoscillations (even if it becomes impractical to use), we find that integrating over the range $[-\arctan\tfrac{1}{\sqrt{a}}, \arctan\tfrac{1}{\sqrt{a}}]$  gives an intensity of

\be
I_{\mathrm{SO}} \approx \frac{2a^{N-1/2}}{2N+1},
\label{Iso_tot_text}
\ee
exponentially smaller than Eq.~\eqref{Itot_approx_sl} by $a^N$.

Let us now consider the SG region characterized by Eq.~\eqref{ylyhtxt}.  We find, for $a\gg 1$, $\kappa$ has the expansion around $x_{sg}$ of $\kappa \approx (aN/2) ( 1 - (a^2/2) (x-x_{sg})^2)$ to quadratic order.  Consequently, we get considerable supergrowth  of approximately exponential form in the region $x \in [x_{sg} - 1/a, x_{sg} + 1/a]$.  Despite the fact that the phase of $f$ oscillates several times in this region, it is of no concern to us, since we are using only the magnitude (intensity) for the imaging.  

Even better exponential fits can be obtained by reducing the range by (say) a factor of 2, but at the cost of reducing the amount of light that is used. The intensity in this reduced region is 
\begin{equation}
\begin{split}
    & I_{\mathrm {SG,R}}\approx \frac{1}{a}\left(1-\tfrac{1}{a^2}\right)^{N-1}\left(\sum_{m=0}^N \binom{N}{m}\frac{2^{2m+1}}{2m+1}-\frac{5^N}{a^2}\right),
\end{split}
   \label{isgr_approx}
\end{equation}
which to order $\tfrac{1}{a}$ is, 
\begin{equation}
\begin{split}
    & I_{\mathrm {SG,R}}\approx \sum_{m=0}^N \binom{N}{m}\frac{2^{2m+1}}{2m+1}\frac{1}{a}.
\end{split}
   \label{Isg_restr}
\end{equation}
It is interesting to note that the intensities in restricted regions Eqs.~\eqref{Iso_restr} and \eqref{Isg_restr} both have $1/a$ dependence, instead of the exponential dependence of total intensities in Eqs.~\eqref{Itot_approx} and \eqref{Iso_tot_text}. This is due to the shrinking of the restricted regions as $a$ is increased.  The coefficient of $1/a$ in the Eq.~\eqref{Isg_restr} is much larger than the corresponding coefficient in Eq.~\eqref{Iso_restr}. Therefore, the near maxima SG region contains more intensity compared to the near maxima SO region.
\begin{figure}
    \centering
    \includegraphics[width=\linewidth]{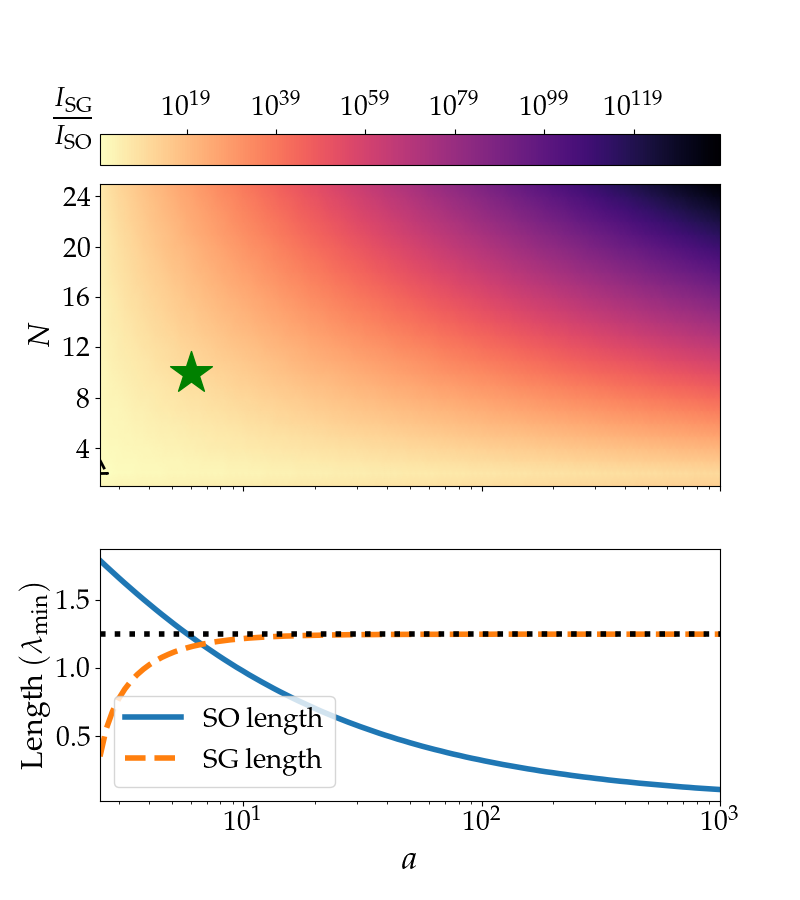}
    \caption{(Top) Ratio of intensities of total SG and SO regions for different values of $a$ and $N$. With the increase of either $a$ or $N$, the intensity within the SG region becomes increasingly higher. The dashed black line on the bottom left corner shows the parameter values with equal SO and SG intensities. The green asterisk identifies the region of parameters chosen for object reconstruction scheme presented in Sec.~\ref{so-sg-imaging}. (Bottom) Comparison of the lengths of SO and SG regions as a function of $a$. While the SO region shrinks with the amount of superoscillation (i.e.~$a$), the   SG length approaches $1.25\lambda_{\min}$. Note that the full range is $N$ independent.
\label{fig:reg_comp}}
\end{figure} 
The intensity in the total SG region ($\kappa>1$ and specified by Eq.~\eqref{ylyhtxt}) can be approximated as 
 \be
I_{\mathrm{SG}}\approx \frac{\pi}{4}\left(\frac{1}{\sqrt{2}}a\right)^{2N} ,
\label{Isg_2text}
\ee
We see that the intensity of the SG region is exponentially higher than that of the SO region by $a^N$, but exponentially smaller than Eq.~\eqref{Itot_approx_sl} by $\left(1/2\right)^{N}$.  Additionally, the great advantage is that we can profitably use the entire SG range, but only a small fraction of the SO range.
This is because the amplitude varies too much once $x$ leaves the more restricted range, as well as the fact that lower values of $k$ become mixed in with the high values (even if still above $N$).  The SG region has none of those drawbacks.


 In Fig.~\ref{fig:reg_comp}, we show quantitative comparisons of SO and SG properties of our chosen function for different values of the parameters $a$ and $N$. We see that SG intensity dominates SO intensity for most of the parameter range, except for a very small region in the bottom left of the plot.  This signifies that for supergrowth imaging can provide a sufficient intensity to obtain bright images, overcoming signal-to-noise ratio issues of superoscillation. We also can compare the lengths of the SO and SG regions with varying $a$ in the bottom panel of Fig.~\ref{fig:reg_comp}. Both of these lengths are independent of $N$. For large values of $a$, the length of the SO region $\propto 2a^{-1/2}$ and approaches 0 asymptotically. However, the length of the SG region $\propto \pi/4 - 2a^{-2}$ and rapidly approaches $\pi/4 = 1.25 \lambda_{\rm min}$ asymptotically for large $a$, which can also be seen in the bottom panel of Fig.~\ref{fns}.

\section{\label{so-sg-imaging} Superresolution  imaging and object reconstruction}

\begin{figure*}
    \centering
\includegraphics[width=0.8\linewidth]{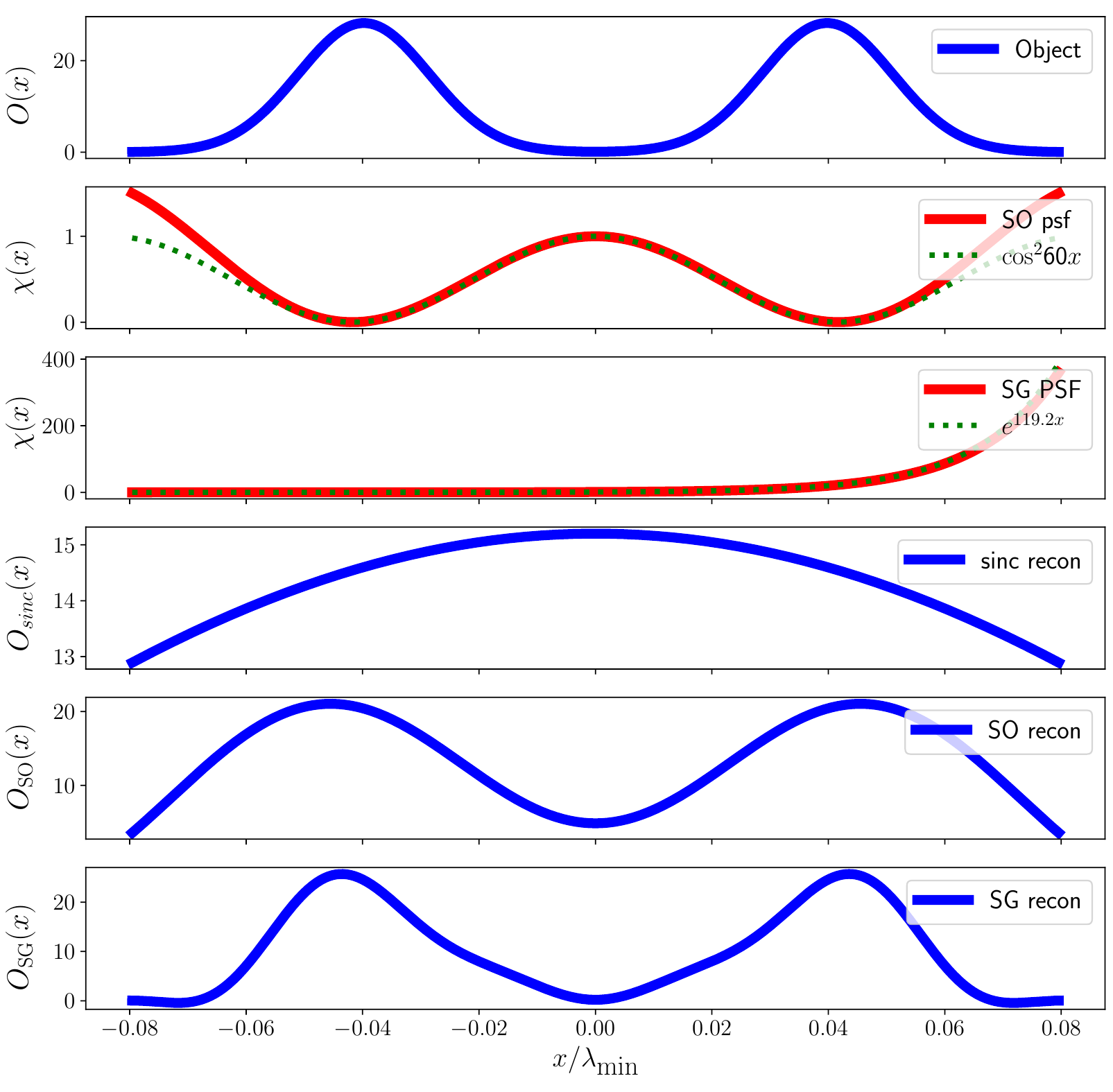}
\caption{Reconstruction of two sub-wavelength peaks using the methods in Sec.~\ref{so-sg-imaging}. The first panel shows the object, which we assume to be two Gaussian peaks  separated by a distance of $\sim0.08\lambda_{\min}$. Throughout the simulations, we assume $N=10$.  The second panel shows the PSF $(\textrm{Re}f(x))^2$ for $a=6$ (solid red) and its approximation (green dashed) $\cos^260x$ near the origin. The third panel shows the PSF  $|f(x)|^2$   (solid red) and its approximation (green dashed) $e^{119.2x}$ centered at $x=x_{sg}$ for $a=12$.  The fourth panel shows object reconstruction with a bandlimited $sinc$ PSF using Eq.~\eqref{ftprod}. The fourth and fifth panels show the results from reconstruction using the methods laid out in Secs.~\ref{so_recon} and \ref{sg_recon} respectively. We see the superior sub-wavelength feature detection ability of our reconstruction scheme.}
    \label{fig:recon_peaks}
\end{figure*}

 In this section, we present schemes for the detection of sub-wavelength features using SO and SG spots in a diffraction limited optical system. For a simple one-dimensional model of incoherent imaging, we define the function \(O(x)\) for the object intensity, and \(S(u)\) for the image intensity.  They are related via the intensity PSF of the imaging system, \(\chi(x)\), which indicates the image created from a point source. Imaging theory dictates
\begin{equation}
\label{eq:img-conv}
S(u) = \int dx O(x) \chi(u-x),
\end{equation}
so the image created is a convolution of the object with the PSF \cite{bornPrinciplesOpticsElectromagnetic1999,goodmanIntroductionFourierOptics2017}.

We consider the problem of reconstructing $O(x)$ from experimental imaging data $S(u)$ \cite{Barnes:66}. Now, the Fourier transforms\footnote{Here adopt the convention that the  Fourier transform of a function $g(x)$ is $\tilde{g}(k)=\int_{-\infty}^{\infty}\tfrac{dx}{\sqrt{2\pi}}e^{-ik x}g(x)$, such that the inverse transform is $g(x)=\int_{-k_{\textrm{max}}}^{k_{\textrm{max}}}\tfrac{dk}{\sqrt{2\pi}}e^{ikx}\tilde{g}(k)$.} of the above functions are related by  
\begin{equation}
\tilde{S}(k) = \tilde{O}(k) \tilde{\chi}(k).
\label{ftprod}
\end{equation}
If the imaging system is illuminated by light of wavelength $\lambda_0=2\pi/k_0$ and has an exit pupil numerical aperture of NA, the PSF and therefore $\tilde{S}(k)$ in Eq.~\eqref{ftprod} is bandlimited by $k_0$NA. Inverting the above equation $\tilde{O}(k)=\tilde{S}(k)/\tilde{\chi}(k)$ leads to a loss of information of the higher spatial frequency components and places a constraint on our ability to resolve features smaller than $\lambda_0$. 

However, in the following investigation we show how isolating the SO/SG region of the function in Eq.~\eqref{f1} and using that as the PSF can help us resolve sub-wavelength features. This could be done in a confocal imaging setup, where the object is illuminated by a  SO/SG optical field spot. 
As we will see, the nature of SO/SG field spot dictates the near origin features of the image, thereby providing an access to the subwavelength features in the object. 
In a confocal setup with one objective, the objective NA and wavelength will determine the extent of the spatial filter function that results from the aperture in front of the detector. A pinhole with a deeply subwavelength diameter will still map to a diameter of $\sim\frac{\lambda}{NA}$ on the object. NA mismatch between illumination and collection is one way to allow for narrow filter functions with respect to SO/SG regions.

In the following analysis, we assume the object has an extent $x \in [-l/2,l/2]$ with $l<\lambda_0$. We can also scan an extended object gradually by filtering  length $l$ at a time.   For further simplicity, we assume $\textrm{NA}=1$ and the object is symmetric (i.e., $O(x)=O(-x)$). Also, in the sample cases we consider, the amplitude PSFs are bandlimited by $N$.  The subsequent analysis separately shows how SO or SG based imaging can help us reconstruct $O(x)$. We assume, for SO based imaging, the intensity PSF is $(\textrm{Re}f(x))^2$ while for SG based imaging the relevant PSF is $|f(x)|^2$.
\subsection{\label{so_recon} Object reconstruction with SO spots}
We consider the real part of Eq.~\eqref{f1}  as the PSF superoscillating at origin with local rate $k=aN\gg N$. Thus near the origin, for a length larger than the object length $l$, $\chi(u-x)\sim\cos^2k(u-x)$ is a good approximation for the PSF and Eq.~\eqref{eq:img-conv} can be written as 
\begin{equation}
\begin{split}
   & S_k(u) = \int_{-l/2}^{l/2} dx O(x) \cos^2k(u-x)\\
   & = C+ \sqrt{\tfrac{\pi}{2}}\tilde{O}(2k)\cos 2ku,
\end{split}
\label{img_SO}
\end{equation}
where for symmetrical objects $\tilde{O}(2k)=\tilde{O}(-2k)$ and $C$ is a constant. The subscript $k$ on $S$   expresses the fact that the observed image intensity depends the chosen PSF.  We can invert Eq.~\eqref{img_SO} or measure the image intensity peak reduction to calculate $\tilde{O}(2k)$.
Thus, Eq.~\eqref{img_SO} provides a prescription for the direct measurement of the high spatial frequency Fourier  coefficients of the object.  We can perform a series of measurements with different values of $k$, i.e.~a series of different PSFs, to map out the Fourier transform of the object. If $k_\textrm{SO}$ denotes the highest value of the local wavenumber in experiment, the reconstructed object is
\begin{equation}
    O_\textrm{SO}(x)=\int_{-2k_{\textrm{SO}}}^{2k_{\textrm{SO}}}\tfrac{dk}{\sqrt{2\pi}}e^{ikx}\tilde{O}(k).
\end{equation}

\subsection{\label{sg_recon} Object reconstruction with SG spots}
In this case, we consider the function in Eq.~\eqref{f1} shifted by $x_{sg}$ such that near origin the PSF can be approximated as $\chi(u-x)\sim e^{\kappa(u-x)}$, where $\kappa=N(a-1/a)\gg N$. Then, we have,
\begin{equation}
   S_\kappa(u) = \int_{-l/2}^{l/2} dx O(x) e^{\kappa(u-x)}.
\label{img_SG}
\end{equation}
Assuming a Fourier expansion of $O(x)$ in $x \in [-l/2,l/2]$, 
\begin{equation}
    O(x)=c_0+\sum_{n=1}^\infty c_n 
    \cos\left(\tfrac{2\pi n x}{l}\right),
    \label{o_fr}
\end{equation}
Eq.~\eqref{img_SG} transforms at $u=0$ to 
\begin{equation}
   S_\kappa(0) = 2\sinh\left(\tfrac{\kappa l}{2}\right) \left[\tfrac{c_0}{\kappa}+\sum_{n=1}^\infty (-1)^nc_n \frac{\kappa}{\kappa^2+\left(\tfrac{2\pi n x}{l}\right)^2}\right].
\label{eqn_SG}
\end{equation}
Note as $n$ increases, the contributions from higher order terms decrease. In practice  the sum on the right hand side will be a good estimate for the left hand side if we  consider terms till $n^\star=\kappa l$. Again we consider a series of different SG PSFs to measure the the left hand side of Eq.~\eqref{eqn_SG} as a function of $\kappa$. We can then solve for the coefficients numerically to reconstruct the object $O_{\textrm{SG}}(x)$.

An alternative approach would be to approximate the image intensity in Eq.~\eqref{img_SG} as a Laplace transform of the object. By performing intensity measurements for different supergrowing PSFs, we gain access to the Laplace transform at different values of $\kappa$. The object then can be reconstructed by performing a numerical inverse Laplace transform  \cite{Linverse}.

One drawback of these schemes is the need to perform imaging using multiple PSFs. In practice, we can limit our experiment to a few different PSFs and use interpolation to approximate $\tilde{O}$ for the intermediate values.

Interestingly, the SO based reconstruction relies on intensity in a small region near origin. Therefore, the method's effectiveness heavily depends on the detector's ability to register small changes in the intensity. On the other hand, the SG based approach does not have this limitation because we only look at the image intensity at a particular point. In experiments, of course,  both of these methods will be limited by noise. Having more intensity, SG gives much better SNR. 

We show the effectiveness of the schemes in Fig.~\ref{fig:recon_peaks}. The top panel shows a sub-wavelength object which is reconstructed using a bandlimited $sinc$ function and inverting Eq.~\eqref{ftprod}, and also using SO/SG based schemes described in this section. The second and third panel compare the PSFs for SO and SG imaging for highest values of $a$ (6 and 12 respectively) with their approximation. As we have already seen in Fig.~\ref{fig:reg_comp}, the length of the SO region shrinks with increasing $a$. This restricts the values of $a$ we can use in practice for SO based reconstruction, while the SG based approach does not have this limitation.    In the next panel we immediately see that the cardinal sine function is unable to resolve the two peaks in the reconstructed object as expected.  However, as the last two panels show,  both the SO and SG PSFs are able to reconstruct the peaks. In the SO based approach, the results of reconstruction are limited by our ability to generate long  superoscillatory features for higher values of $a$. Whereas, in SG based approach, the numerical accuracy in determining the coefficients $c_n$ in Eq.~\eqref{o_fr} limits the reconstruction. These shortcomings in both cases lead to slightly  shifted estimation of the  peaks.

\section{\label{disc} Discussion}
It was previously shown~\cite{jordanSuperresolutionUsingSupergrowth2020} that the function in Eq.~\eqref{f1} is capable of superresolution imaging using the superoscillation and supergrowth using the function over a full period. For the case of two nearby point sources, putting the point sources in the ``sweet spot'' of the SO or SG region (maximum $k(x)$ or $\kappa(x)$ respectively) can give superresolution of the two sources. However, due to the nature of the function, any apparatus aiming to use this method would require extremely high dynamic range, both on the illumination and detection side, which limits its utility. We overcome this limitation on the detection side by 
applying spatial filtering of sweet spot and scanning the image.
While both behaviors can resolve features better than the Fourier bound, the SG phenomena benefits from exponentially higher intensity of the illumination point spread function, whereas using SO behavior would require incredibly sensitive detectors due to the exponentially suppressed amplitude. There is also the concern of hitting the diffraction limit for the SO PSF, due to the shrinking length of the region as the parameter \(a\) is increased, which is not an issue for the SG case. This demonstrates the superiority of SG spots as a means to achieve superresolution.

\section{\label{concl} Conclusions and outlook}
We describe supergrowth as a concept analogous to superoscillations, except we evaluate the local growth rate instead of the local wave number of a function. We characterize the supergrowing and superoscillating regions of a canonical oscillatory function as well as  provide analytical approximations for the energy inside total supergrowing, total superoscillating, near maxima supergrowing and near maxima superoscillating regions. Our analysis reveals that the supergrowing regions can contain intensity that is exponentially larger in terms of the highest local wavenumber compared to that of the superoscillating regions.  These results indicate that superresolution imaging using supergrowing spots could be more advantageous compared to superoscillation based superresolution imaging.  Finally, we numerically show  that the SO and SG based superresolution imaging is able to reconstruct objects beyond the diffraction limit, thereby demonstrating the efficacy of our schemes.

These findings highlight supergrowth as a potentially superior far-field superresolution scheme. Physically, superoscillation corresponds to rapidly oscillating regions with small amplitude. However, supergrowth, due to high growth rate, connects the smaller amplitude regions with higher amplitude ones. Therefore, it is not surprising that   supergrowing regions can contain significantly more light while also giving enhanced resolution.  The analytical approximations for intensities characterize the parametric dependence of the light within different regions. These expressions can be useful for designing optimized superoscillatory or supergrowing lenses. Furthermore, the object reconstruction schemes  provide the framework for experimental implementation of supergrowth and superoscillation  based superresolution imaging.

Our results offer a variety of avenues for potential future investigations. The analysis presented here is restricted to a simple 1-d oscillatory function. Performing  similar calculations for more general and higher dimensional functions could have greater significance for experimental implementation of supergrowth imaging. On that note, the natural next step is to investigate ways to generate optimized supergrowing spots.   Another relevant problem is to explore whether  simultaneous occurrence of superoscillations and supergrowth could be leveraged for better resolution with suppressed side lobe intensity.  Apart from the prospective theoretical ventures, a lab realization of supergrowth based superresolution imaging is an imminent experimental challenge. Therefore, our analysis serves as the groundwork for a novel superresolution and object reconstruction scheme    with a multitude of  scopes for further prospects.

\section*{\label{sec:level1}Acknowledgement}
We thank  Sethuraj Karimparambil Raju, Sultan Abdul Wadood, Anurag Sahay for providing insight throughout the project. This work has been supported by the AFOSR grant \#FA9550-21-1-0322 and the Bill Hannon Foundation.  
 \appendix
 
\section{\label{app0}Intensity calculation}

The total intensity \eqref{I0} can be written as 
\be
I_0=2\int_{-\tfrac{\pi}{2}}^{\tfrac{\pi}{2}} (\cos^2{x}+a^2\sin^2{x})^N dx
\label{I0_1}
\ee
A change of variables $y=\tan{x}$ leads to 
\be
I_0=2\int_{-\infty}^{\infty} \frac{(1+a^2y^2)^N}{(1+y^2)^{N+1}} dy
\label{I0_2}
\ee
Performing a binomial expansion of the numerator and also noting that the integrand is an even function, we can write
\be
I_0=4\sum_{m=0}^N\binom{N}{m}\int_{0}^{\infty} \frac{a^{2m}y^{2m}}{(1+y^2)^{N+1}} dy.
\label{I0_3}
\ee
Adopting another change of variables $y^2=\tfrac{z}{1-z}$, the integral $\int_{0}^{\infty} \frac{y^{2m}}{(1+y^2)^{N+1}} dy$  can be evaluated in terms of Beta functions to be $\tfrac{1}{2}\mathrm{B}(m+\tfrac{1}{2},N-m+\tfrac{1}{2})$. Therefore, using the relationship between Beta and Gamma functions, 
\be
I_0=2\sum_{m=0}^N\binom{N}{m}\frac{\Gamma(m+\tfrac{1}{2})\Gamma(N-m+\tfrac{1}{2})}{\Gamma(N+1)} a^{2m}.
\label{I0_4}
\ee
For large $a$ and $N$, we only consider the highest order term in $a$ i.e.~$a^{2N}$. Thus the approximate intensity is
\be
I_0\approx2\pi\frac{(2N)!}{(N!)^2}\left(\frac{a}{2}\right)^{2N}.
\label{I0_f_1}
\ee

Near the fastest oscillation region $x \in [-(N(a^2-1))^{-1/2},(N(a^2-1))^{-1/2}]$, the intensity is 
\be
I_{\mathrm {SO,R}}=2\int_{0}^{\frac{1}{\sqrt{N(a^2-1)}}}(\cos^2{x}+a^2\sin^2{x})^N dx.
\label{Iso_1}
\ee
This can be approximated as
\be
I_{\mathrm {SO,R}}\approx 2\int_{0}^{\frac{1}{\sqrt{N(a^2-1)}}}\exp\{N(a^2-1)x^2\}  dx.
\label{Iso_2}
\ee
Now for large $a$ and $N$, $\frac{1}{\sqrt{N(a^2-1)}}\ll1$
. Using the approximate mean value theorem for integrals, the integration leads to
\be
I_{\mathrm {SO,R}}\approx\frac{2e^{1/4}}{\sqrt{N(a^2-1)}}.
\label{Iso_3}
\ee
The function behaves like $\sim e^{iaN}$ in this restricted region. If we only consider the real part of $f(x)$, i.e.~if our PSF is $(\textrm{Re}f(x))^2$, then the corresponding   intensity is almost half of Eq.~\eqref{Iso_3}. Thus, the intensity pertaining to the real part can be approximated as
\be
I_{\mathrm {SO,R}}^{\textrm{Re}}\approx\frac{e^{1/4}}{\sqrt{N(a^2-1)}}.
\label{Iso_re_3}
\ee

Next, we look at the total intensity in the superoscillatory spot  $x \in [-\arctan\tfrac{1}{\sqrt{a}},\arctan\tfrac{1}{\sqrt{a}}]$, i.e.,
\begin{equation}
    I_{\mathrm{SO}}=2\int_{0}^{\tan^{-1}\tfrac{1}{\sqrt{a}}}(\cos^2 x+a^2 \sin^2 x)^N dx.
    \label{Iso_expr}
\end{equation}
Assuming $y=\tan x$, the above integral can be transformed as 
\be
I_{\mathrm{SO}}=2\int_{0}^{1/\sqrt{a}}\frac{(1+a^2y^2)^N}{(1+y^2)^{N+1}} dy,
\label{Iso_a}
\ee
with $y\ll 1$ for large $a$. Using binomial expansion, we can write
\be
\begin{split}
&I_{\mathrm{SO}}=2\int_{0}^{1/\sqrt{a}}dy\bigg[\sum_{m=0}^{N}{N \choose m}a^{2m}y^{2m}\times\\&\left(1-(N+1)y^2+\frac{(N+1)(N+2)}{2!}y^4+\dots\right)\bigg]\\=& 4\bigg[\sum_{m=0}^{N}{N \choose m}\left(\frac{a^{m-1/2}}{2m+1}-\frac{(N+1)a^{m-3/2}}{2m+3}+\dots\right)\bigg].
\end{split}
\label{Isoch_2}
\ee
Taking only the highest order term in $a$, we get
\begin{equation}
I_{\mathrm {SO}}\approx\frac{2a^{N-1/2}}{2N+1}.
    \label{Isoch_f}
\end{equation}
If we only consider the real part of $f(x)$, the intensity within is bounded by Eq.~\eqref{Isoch_f}. Therefore, $I_{\mathrm {SO}}^{\textrm{Re}}\leq I_{\mathrm {SO}}$. 

Next we calculate the intensities corresponding to the supergrowing regions. From Eq.~\eqref{kappa}, we can write $\kappa/N$ in terms of $y=\tan{x}$
\be
\frac{\kappa}{N}=\frac{(a^2-1)y}{1+a^2y^2}.
\label{kappa_1}
\ee
The region of supergrowth corresponds to $y_l<y<y_h$, where

\begin{equation}
    \begin{split}
        &y_l=\tfrac{1}{2}\left(1-\tfrac{1}{a^2}-\sqrt{(1-\tfrac{1}{a^2})^2-\tfrac{4}{a^2}}\right),\\
        &y_h=\tfrac{1}{2}\left(1-\tfrac{1}{a^2}+\sqrt{(1-\tfrac{1}{a^2})^2-\tfrac{4}{a^2}}\right).
    \end{split}
    \label{ylyh}
\end{equation}
Similarly, the superdecaying region corresponds to $-y_h<y<-y_l$.

Now, we consider the near extrema ($x_{sg}=\arctan{\tfrac{1}{a}}$) region for the supergrowth. The point of extrema can be written as $\tan^{-1}{\tfrac{1}{a}}=\left({\tfrac{1}{a}}-{\tfrac{1}{3a^3}}+{\tfrac{1}{5a^5}}-\dots\right)$. Therefore, $x\in [x_{sg}-1/a,x_{sg}+1/a]$ corresponds to the total region of $\sim[-\tfrac{1}{3a^3},\tfrac{2}{a}-\tfrac{1}{3a^3}]$. However, since $\arctan y_l>0$, the region of interest should instead be $\sim\left[ \arctan y_l, x_{sg}+1/a\right]$. Thus,
\be
\begin{split}
I_{\mathrm {SG,R}}=
     \int_{\arctan y_l}^{x_{sg}+\tfrac{1}{a}}\left(\cos^2 x+a^2\sin^2x \right)^Ndx.
\end{split}
\label{Isgr}
\ee
With the change in variable $x^\prime=x-x_{sg}$, we can write 
\be
\begin{split}
& I_{\mathrm {SG,R}}=\left(\frac{2a^2}{1+a^2}\right)^N\times\\&
     \int_{\arctan y_l-\arctan\tfrac{1}{a}}^{\tfrac{1}{a}}\left(1+\rho \sin x^\prime\cos x^\prime +\tfrac{1}{2}\rho^2\sin^2x^\prime \right)^Ndx^\prime,
\end{split}
\label{Isgr_1}
\ee
where $\rho=a-\tfrac{1}{a}$. For large $a$, the interval could be approximated as $\left[-\tfrac{1}{a},\tfrac{1}{a}\right]$, and the integral is approximately

\begin{equation}
    I_{\mathrm {SG,R}}\approx \left(\frac{2a^2}{1+a^2}\right)^N \int_{-\tfrac{1}{a}}^{\tfrac{1}{a}} \left(1+\rho x^\prime +\tfrac{1}{2}\rho^2 x^{\prime 2} \right)^Ndx^\prime.
    \label{isgr_approx1}
\end{equation}


Performing the integral leads to \begin{equation}
\begin{split}
    & I_{\mathrm {SG,R}}\approx \left(\frac{a^2}{1+a^2}\right)^N \times\\&\sum_{m=0}^{N}\frac{1}{\rho }\binom{N}{m}\frac{(1+\rho/a)^{2m+1}-(1-\rho/a)^{2m+1}}{2m+1}.
\end{split}
    \label{isgr_approx2}
\end{equation}
For $a,N\gg 1$, this could be further approximated to
\begin{equation}
\begin{split}
    & I_{\mathrm {SG,R}}\approx \tfrac{1}{a}\left(1-\tfrac{1}{a^2}\right)^{N-1}\left(\sum_{m=0}^N \binom{N}{m}\frac{2^{2m+1}}{2m+1}-\frac{5^N}{a^2}\right).
\end{split}
   \label{isgr_approx3}
\end{equation}
The intensity first increases, then decreases as a function of $a$. This is because at  the value of the function $f(x)$ increases with $a$. However, as we keep increasing $a$, the length of the restricted region gets smaller leading to a decrease in the intensity. The approximate value of $a$ with maximum $ I_{\mathrm {SG,R}}$, denoted as $a^\star$, could be calculated by differentiating the above expression. Assuming, $C_N=\sum_{m=0}^N \binom{N}{m}\frac{2^{2m+1}}{2m+1}$, $\gamma_N=5^N/C_N$ and $\beta_N=\tfrac{1}{2}\left(2N-1+3\gamma_N\right)$, we get
\begin{equation}
    a^\star(N)\approx\sqrt{\sqrt{\beta_N^2-\gamma_N}+\beta_N}.
\end{equation}

Lastly, we consider the total intensity in the supergrowing region
\be
I_{\mathrm{SG}}= \int_{\arctan y_l}^{\arctan y_h}\left(\cos^2 x+a^2\sin^2x \right)^Ndx.
\label{Isg_tot}
\ee
 For large $a$, the upper and lower bounds can be approximated to be $0$ and $\pi/4$ respectively. Similar to Eq.~\eqref{I0_3} the integration above can written as  
\be
I_{\mathrm{SG}}\approx\sum_{m=0}^N\binom{N}{m}\int_{0}^{1} \frac{a^{2m}y^{2m}}{(1+y^2)^{N+1}} dy.
\label{Isg_1}
\ee
 Only limiting to highest order term in $a$ leads to 
 \be
I_{\mathrm{SG}}\approx F_N a^{2N},
\label{Isg_2}
\ee
where
\be
F_N=\int_0^1 dy \frac{y^{2N}}{(1+y^2)^{N+1}},
\label{Fneqn}
\ee
 can be evaluated using the recurrence relation 
\be
F_N = (1-\tfrac{1}{2N})F_{N-1}-\frac{1}{2^{N+1}N},
\ee
and $F_0=\pi/4$. We can find the asymptotic value of $F_N$ by transforming  the integral in Eq.~\eqref{Fneqn} with $y=\tan\theta$. Then, $F_N=\int_0^{\pi/4}d\theta\sin^{2N}\theta$. The integral can be expressed in terms of incomplete Beta function as $F_N=\tfrac{1}{2}B_{\tfrac{1}{2}}(N+\tfrac{1}{2},\tfrac{1}{2})$. As $N$ increases, the integrand approaches zero. We approximate $\sin\theta\approx\frac{1}{\sqrt{2}}$ in the interval $\theta\in[0,\pi/4]$. Thus, $F_N\sim \left(1/2\right)^{N}\frac{\pi}{4}$. 
\typeout{}
\bibliography{refs}
\end{document}